\documentclass[%
 aip,
 amsmath,amssymb,
 reprint,%
]{revtex4-2}

\usepackage{graphicx}
\usepackage{dcolumn}
\usepackage{bm}

\usepackage[utf8]{inputenc}
\usepackage[T1]{fontenc}
\usepackage{mathptmx}
\usepackage{etoolbox}
\usepackage{float}

\makeatletter
\def\@email#1#2{%
 \endgroup
 \patchcmd{\titleblock@produce}
  {\frontmatter@RRAPformat}
  {\frontmatter@RRAPformat{\produce@RRAP{*#1\href{mailto:#2}{#2}}}\frontmatter@RRAPformat}
  {}{}
}%
\makeatother
\begin{document}
\title{Integration of silicon-vacancy centers in nanodiamonds with an optical nanofiber}
\author{Ramachandrarao Yalla*}
\email{rrysp@uohyd.ac.in}
\affiliation{School of Physics, University of Hyderabad, Hyderabad, Telangana, India-500046}
\author{Y. Kojima}
\author{Y. Fukumoto}
\author{H. Suzuki}
\author{O. Ariyada}
\affiliation{Arios Co., Ltd., Akishima, Tokyo, 196-0021, Japan}
\author{K. Muhammed Shafi}
\altaffiliation{Department of Instrumentation \& Applied Physics, Indian Institute of Science, Bengaluru, India-560012}
\author{Kali P. Nayak}
\altaffiliation{Department of Engineering Science, University of Electro-Communications, Chofu, Tokyo, Japan-182-8585}
\author{Kohzo Hakuta}
\affiliation{Center for Photonic Innovations, University of Electro-Communications, Chofu, Tokyo, Japan-182-8585}
\date{\today}

\begin{abstract}
We experimentally demonstrate the integration of silicon-vacancy centers in nanodiamonds (SiV-NDs) with an optical nanofiber (ONF). We grow SiV-NDs on seed NDs dispersed on a quartz substrate using a microwave plasma-assisted chemical vapor deposition method. First, we search and characterize SiV-NDs on a quartz substrate using an inverted confocal microscope and an atomic force microscope (AFM). Subsequently, we pick up SiV-NDs from the quartz substrate and deposit them on the surface of a free-standing ONF using the AFM tip. The fluorescence emission spectrum, photon count rate, and intensity correlations for SiV-NDs are systematically measured. 
\end{abstract}

\maketitle
Integrating quantum emitters with nanophotonic structures has attracted great interest towards potential applications in quantum photonics \cite{Sipahigil,Lodahl}. Various techniques have
been proposed and demonstrated for the direct fabrication of nanophotonic structures
with embedded quantum emitters\cite{Sipahigil,Lodahl, Riedrich,Englund10}. Examples include microcavities\cite{Lodahl}, photonic crystal cavities\cite{Riedrich,Englund10}, diamond nanobeams\cite{Sipahigil}, and nanowaveguides \cite{Lodahl}. In contrast, a hybrid approach has been received a growing platform, wherein quantum emitters are integrated with photonic nanostructures through a bottom-up technique for precise and flexible positioning\cite{Lodahl,Review18}.  

Regarding quantum emitters, solid-state quantum emitters such as quantum dots, defect centers in nanodiamonds (NDs), and molecules would be preferable for quantum photonic applications\cite{Kurtsiefer,Aharonovichrev}. Among solid-state quantum emitters, defect centers in NDs have been proven to be a potential candidate for practical single-photon applications\cite{Ladd,Aharonovich11,Englund}. Various defect centers in NDs have been proposed and developed in recent years\cite{Rabeau,Aharonovich11}. To date, experiments have been performed extensively on nitrogen-vacancy (NV) centers in NDs (NV-NDs) due to their spin manipulation capability and established fabrication technology. However, NV-NDs emit a few percentages of the whole emission into zero phonon line (ZPL) at room temperatures\cite{Aharonovich11}. Recently, silicon-vacancy centers in NDs (SiV-NDs) have become promising single-photon source candidates due to the following advantages\cite{Aharonovich11,Neu}. (i) A distinct ZPL emission with a narrow spectral width even at room temperatures. (ii) A four-line fine structure of the ZPL at cryogenic temperatures. (iii) Nearly lifetime-limited spectral linewidths with optically accessible spin states. (iv) Less prominent background of the host material around the ZPL emission wavelength, making it easily observable in the experiments. Moreover, SiV-NDs are feasible to synthesize using currently existing technology. Various methods have been proposed and developed for synthesizing SiV-NDs, such as the chemical vapor deposition (CVD) \cite{Neu}, ion implantation \cite{ Meijer, Takashima}, and high-pressure and high-temperature methods\cite{Davydov,Jantzen}. Among them, the CVD method is preferable due to the flexible choice of substrate, precise control of ND size and density, and easy incorporation of any specific dopant\cite{Aharonovich11}. Specifically, the CVD method has been proven to be a potential candidate for synthesizing SiV-NDs as single-photon sources\cite{Neu}. 

Regarding nanophotonics structures, sub-wavelength diameter silica fibers, termed as optical nanofibers (ONFs),  have become a promising avenue in various aspects for practical quantum
photonics applications due to the strong confinement of the optical field along with automatic lossless coupling to conventional single-mode fibers\cite{Kali19,Yalla12, Yalla14}. The integration of solid-state quantum emitters on the surface of the ONF has been achieved using various techniques such as sub-pico litere needle dispenser\cite{Yalla12, Yalla14, Yalla11}, drop-casting\cite{Schroder}, metal tip\cite{Schell15}, and atomic force microscope (AFM)\cite{Liebermeister}. Among those techniques, the AFM technique has been a proven potential candidate due to the capability of positioning solid-state quantum emitters at a required location with high accuracy\cite{Schell11,Liebermeister}. The AFM technique has been applied for the drop-casted NDs on the substrate\cite{Schell11,Liebermeister}, but it has not been applied to CVD-grown NDs on the substrate. Moreover, the integration of SiV-NDs with a free-standing ONF has not been experimentally demonstrated yet. 

In this Letter, we report the integration of  SiV-NDs with a free-standing ONF by combining an inverted confocal microscope (ICM) with the AFM. We grow SiV-NDs on seed NDs dispersed on a quartz substrate using a microwave plasma-assisted CVD method. First, we search and characterize SiV-NDs on a quartz substrate using the ICM and the AFM. Subsequently, we pick up SiV-NDs from the quartz substrate and deposit them on a free-standing ONF using the AFM tip. The fluorescence emission spectrum, photon count rate, and intensity correlations for SiV-NDs are systematically measured.  

\begin{figure*}[ht]
\centering
\includegraphics[width=14.5cm]{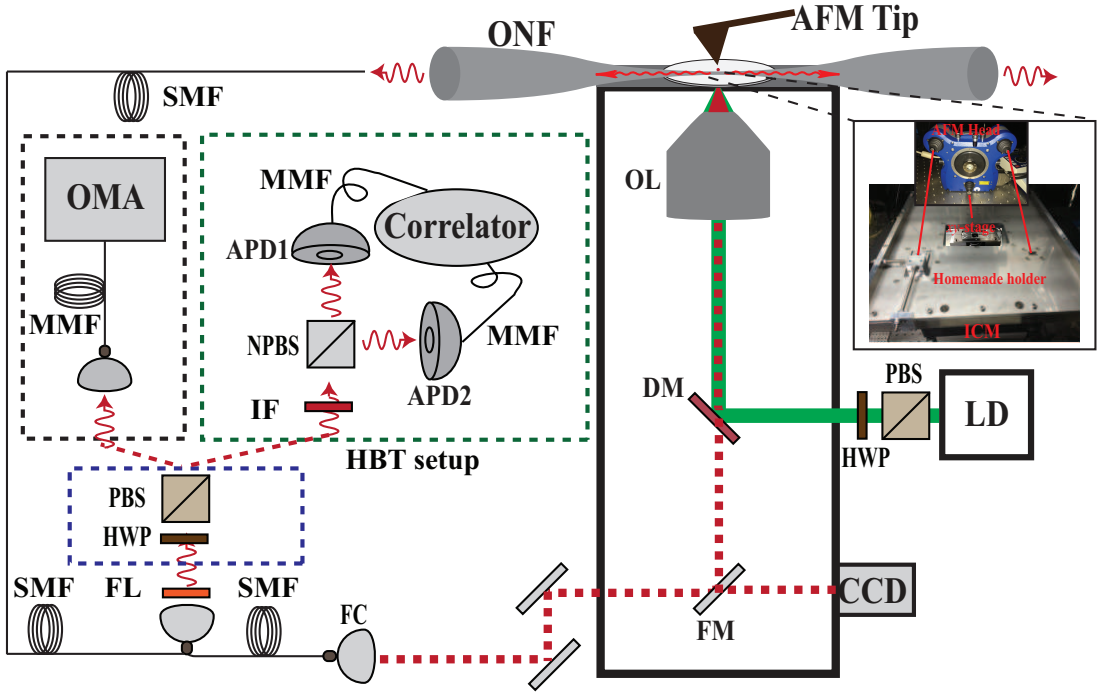}
\caption{{\label{Schematic}}The schematic diagram of the experimental setup. The inverted confocal microscope (ICM) is used to characterize fluorescence emission from SiV-NDs. The inset shows the ICM combined with an AFM tip is used for the pick-up and deposit of SiV-NDs onto the optical nanofiber (ONF). OL, DM, FM, FL, NPBS, PBS, HWP, and IF denote objective lens, dichroic mirror, flipper mirror, filter, non-polarizing beam splitter, polarizing beam splitter, a half-wave plate, and interference filter, respectively. SMF and MMF denote single-mode fiber and multimode fiber, respectively. AFM, FC, APD, LD, and OMA denote atomic force microscope, fiber coupler, avalanche photodiode, laser diode, and optical multichannel analyzer, respectively.} 
\end{figure*}

First, we discuss the fabrication of ONFs and the synthesis of SiV-NDs samples for the experiments.    ONFs are fabricated from single-mode fibers (Fujikura, FutureGuide-SM) using the heat and pull technique with a diameter of $530$ nm, uniform waist length of  $2.5$ mm, and optical transmission of $96$\% at a wavelength of 760 nm\cite{Review18}. Regarding SiV-NDs synthesis, a microwave-plasma-assisted CVD growth method is employed on seed NDs dispersed on a quartz substrate. We use commercially available high purity aqueous synthetic NDs (Microdiamant Liquid Diamond, MSY 0-0.03 GAF) as seed with sizes up to 30 nm. The diluted ND-solution is centrifuged (Chibation-II, XX42CF00T) for 20 min with a speed of 10 krpm; subsequently, the solution was fully de-agglomerated for 30 min using an ultra Sonifier (Branson, SFX250). Such ND-solution is spin-coated onto quartz substrates using a spinner (Kyowariken, K-359SD-2) with a speed of 2500 rpm. We use such NDs deposited substrates for the CVD process. The optimum CVD parameters are as follows. Hydrogen/methane ratio of 100:1, gas pressure of 10 kPa, and a microwave power of 150 W. The CVD process is performed for 15 minutes for currently investigated samples. Due to the presence of silicon atoms originating from quartz substrates themselves, SiV centers are incorporated into NDs during the CVD process. Note that no additional silicon source was injected during the CVD process.
 
Next, we discuss the experimental techniques to characterize the CVD-grown SiV-ND samples. A schematic diagram of the experimental setup is illustrated in Fig.\ref{Schematic}. The essential point is to combine an ICM (Nikon, Eclipse Ti-U) with an AFM (JPK Instruments, NanoWizard). The ICM setup characterizes the fluorescence emission from SiV-NDs. The AFM is used for two purposes. One is to measure the size of SiV-NDs in tapping mode, and the other is to pick up and deposit onto a free-standing ONF in contact mode. For scanning the sample, a high precision (step resolution of 50 nm) $xy$-stage (Sigma Koki, Bios-405T) is installed on top of the ICM as shown in the inset of Fig.\ref{Schematic}. The AFM head is installed onto a homemade holder on the ICM with a three-axis tip scanner. The positioning of the sample and AFM tip can be controlled independently to the focus point of the ICM.

The CVD-grown quartz substrate/ONF is aligned to the focus point of the ICM. SiV-NDs are excited through an objective lens OL (Nikon, 100X, NA= 0.85) at the wavelength of 532 nm derived from a laser diode (LD). The laser spot size at the focus point is 0.6 $\mu$m. Regarding experiments with quartz substrates, the fluorescence collection is performed using the same OL. We coupled the fluorescence light into a single-mode fiber (Thorlabs, 780HP) to realize a confocal pinhole. The fluorescence is separated from reflected laser light using a dichroic mirror (Thorlabs, DMLP650R). Regarding experiments with ONF, the fluorescence is coupled to the guided modes of the ONF and measured at the end of the single-mode fiber. A color glass filter FL (HOYA, O-56) is used to reject the scattered excitation laser light for all-optical measurements. A narrow bandwidth interference filter IF (Semrock, FF01-740/13-25) is used to filter the ZPL emission in the spectral range of 727-752 nm. The IF is used for fluorescence imaging, photon counting, and correlation measurements. 

The ZPL fluorescence is directed to a Hanbury-Brown-Twiss (HBT) setup to measure intensity correlations. The HBT setup employs a non-polarizing beam splitter (NPBS), two avalanche photodiodes (APDs) (Perkin Elmer, SPCM-AQR/FC), and a two-channel single-photon counter (Pico Quant GmbH, Pico-Harp 300). The photon arrival times at both APDs are recorded, and correlations are derived. For spectral characterization, the fluorescence spectrum is recorded using an optical multichannel analyzer (OMA) (Andor, DV420A-OE) with a resolution of 1.5 nm. To measure the polarization properties of SiV-NDs, a half-wave plate (HWP) and the polarizing beam splitter (PBS) are used in the excitation and emission (detection) path. We image the sample/AFM tip through the OL using a charge-coupled device (CCD). 

First, we identify SiV-NDs on quartz substrates by performing confocal fluorescence scans through the microscope OL. The sample is mounted on the $xy$ stage and the fluorescence light intensity is recorded using the APD. The typical fluorescence scan area is 25$\times$25 $\mu m^2$ with a step size of 500 nm. The fluorescence photons are recorded in the time bin of 0.5 s for each $x$ and $y$-position. Note that the system is carefully designed to minimize the position drift during the scanning of the sample. We measure the fluorescence emission spectrum and photon correlations for each SiV-NDs location. Only NDs having SiV centers are used for subsequent pick-up procedures. 

\begin{figure}[ht]
\centering
 \includegraphics[width=9cm]{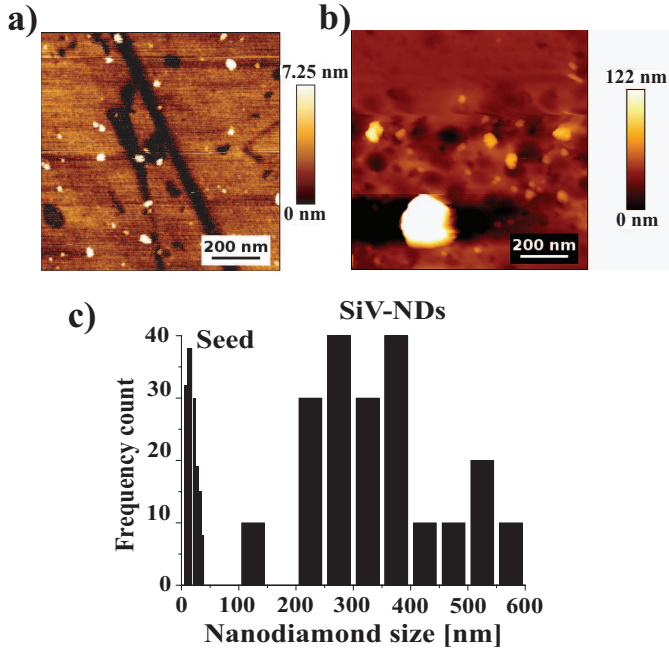}
\caption{{\label{NDs size}} (a) and (b) are typical AFM topography images of seed and SiV-NDs, respectively. (c) A histogram of the size distribution of seed and SiV-NDs for currently investigated samples.} 
\end{figure}

We perform AFM imaging for seed NDs and SiV-NDs samples to measure the size and density distribution. Typical AFM images of seed NDs and SiV-NDs are shown in Figs. \ref{NDs size} (a) and (b),  respectively. A histogram of the measured size distribution of seed and SiV-NDs is shown in Fig. \ref{NDs size} (c). For SiV-NDs (seed NDs), the mean of the size distribution is about 340 nm (18 nm), with a standard deviation of 120 nm (12 nm). The estimated density for SiV-NDs (seed NDs) is roughly a few/$\mu m^2$ (10-12 /$\mu m^2$) for currently investigated samples. Therefore, the percentage of NDs containing SiV centers after the CVD process is less than 10\%. 
 
\begin{figure}[ht]
\centering
\includegraphics[width=9cm]{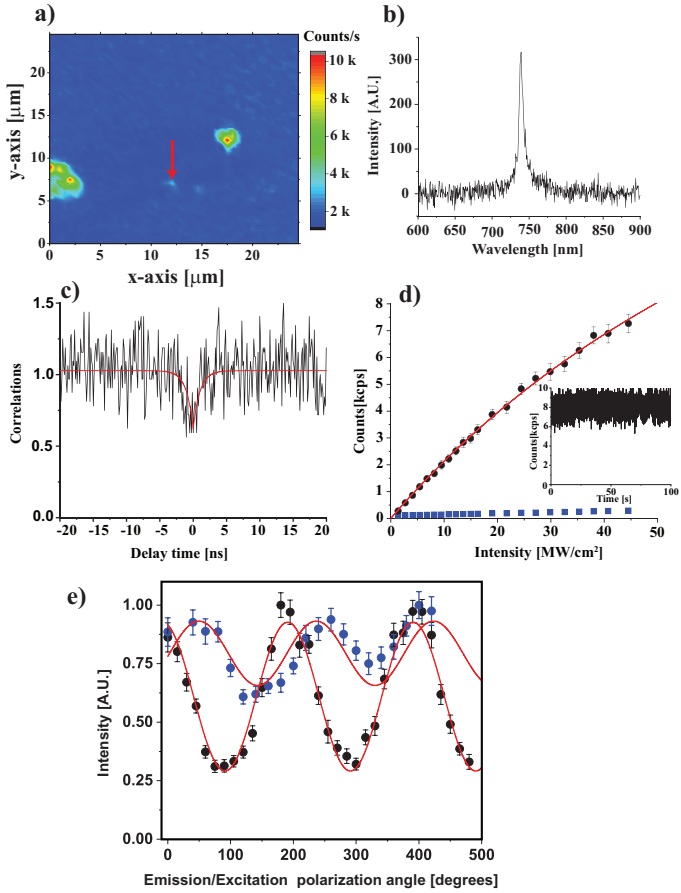}
\caption{{\label{NDs pre-characterization}}(a) A typical confocal fluorescence scan image. (b) The observed SiV-ND fluorescence emission spectrum for the marked position. (c) The normalized intensity correlations versus delay time. (d) Photon-count rate versus excitation laser intensities. Balck circles (signal) and blue squares (background) correspond to the SiV-ND and quartz substrate, respectively. The inset shows the fluorescence count rate as a function of time. (e) Polarization-dependent photon-count rates versus emission/excitation polarization angles.}
\end{figure}

A typical confocal fluorescence scan image is shown in Fig. \ref{NDs pre-characterization}(a). 
Note that our study mainly focused on low photon count rate locations for finding a single or a few SiV-NDs. A detailed characterization of SiV-ND for a marked (red arrow) position is discussed in the following. Figure \ref{NDs pre-characterization} (b) displays a typical fluorescence emission spectrum of SiV-NDs. One can readily see the ZPL emission in the fluorescence spectrum. We fit the fluorescence emission spectrum with the Lorentzian function. The emission spectrum exhibits a peak position wavelength of around 738.8 nm with a full-width at half maximum (FWHM) of 7 nm. The observed ZPL linewidth at room temperature is broader in comparison to previous studies\cite{Neu}. This may be due to the broadening mechanisms related to electron-phonon interaction and local lattice distortions as discussed in Ref\cite{Neu} and resolution of the OMA. The true FWHM should be less than 7 nm after considering the resolution of the OMA. The estimated Debye-Waller factor is about 0.74 for the observed emission spectrum. The measured emission spectrum resembles previously studied SiV-NDs on iridium substrate\cite{Neu}. 

The measured normalized intensity correlations are shown in Fig. \ref{NDs pre-characterization} (c). A signature of the quantum emitter is observed as an antibunching dip at zero delay. We fit the anti-bunching signal by a single exponential function (red curve). The estimated decay time is $1.0 (\pm0.3)$ ns. The estimated decay time is of the same order as previously measured SiV-NDs on iridium substrate\cite{Neu}. The correlation at the anti-bunching dip is $0.6$. The contributions to the dip value are as follows. One is from the background of quartz substrate and host ND itself, and the other is from the timing resolution of the HBT setup. We estimate the expected correlations dip value for the same host ND using the measured signal-to-background ratio from the SiV-NDs emission spectrum as shown in Fig. \ref{NDs pre-characterization} (b). The expected correlation dip value is 0.4 for the measured signal-to-background ratio of 3.5. Therefore, the corrected dip value would be 0.2 after considering background contribution. Additionally, the timing resolution of the current setup is 300 ps, including the response times of APDs. The timing resolution affects the decay time and dip value. Therefore, one can deconvolute the measured correlations with the measured timing response function of the setup. Then, the corrected dip value is less than 0.2, revealing that the number of SiV centers may be one. Also considering the timing resolution, the true decay time should be less than 1 ns.   

We investigate photon-count rates for SiV-ND and quartz substrate at various excitation laser intensities. Figure \ref{NDs pre-characterization} (d) displays a summary of typical photon-count rate versus excitation laser intensities. Balck circles (signal) and blue squares (background) correspond to photon count rates for SiV-ND and quartz substrate, respectively. Note that the signal to background ratio is more than 25. Typical fluorescence photon count rate as a function of time at the excitation laser power of 135 mW (47.7 MW/$cm^2$) is shown in the inset of Fig. \ref{NDs pre-characterization} (d). The background level on the quartz substrate is as low as 0.3 kcps even for the highest excitation powers, implying the benefit of the quartz substrate. The real background should also include the fluorescence of the host ND itself as discussed. Note that the currently investigated SiV-NDs did not show blinking behavior of the photon-count rates as seen in the inset of Fig. \ref{NDs pre-characterization} (d). We fit the observed signal count rate ($n_{obs}(I)$) dependence on the excitation intensity ($I$) with a function $n_{obs}(I)$= $n_{obs}(\infty )$$I$/($I+I_{sat}$), where  $I_{sat}$ and $n_{obs}(\infty )$ are saturation intensity and maximum observable photon-count rate, respectively. The estimated $n_{obs}(\infty )$ and $I_{sat}$-values are 29 kcps and 130 MW/$cm^2$, respectively. This corresponds to a saturation excitation power of 370 mW.

We also measure $n_{obs}(I)$-value dependence on emission/excitation polarization angles, while fixing the excitation/emission polarizer angle at maximum value of $n_{obs}(I)$. The summary of the measured results is shown in Fig.\ref{NDs pre-characterization} (e). Black circles (blue circles) denote emission (excitation) polarization angles. One can readily see the strong emission (excitation) polarization dependence. We fit the measured $n_{obs}(I)$-values with a sine-square function (red curve). We estimate visibility ($V$) to be $0.54$ and $0.25$ for the emission and excitation polarization, respectively. The expected visibility is close to 100\%, assuming that SiV-ND is a single dipole. The orientation of SiV center dipoles in ND is random due to the random orientation of the NDs on the substrate. Therefore, the deviation may be due to polarization anisotropy induced by a dipole orientation with respect to the optical axis of the OL, strong focusing of the beam, the dichroic mirror (DM), and the above band-edge excitation.

Next, we discuss the procedure for picking up and depositing SiV-NDs on the surface of free-standing ONFs. Free-standing is important for avoiding transmission loss due to contamination from the substrate and maintaining high-quality optical properties of ONFs. Deformed AFM tips are used for this purpose. First, SiV-ND location is identified using AFM in tapping mode. We start by positioning the AFM tip above the SiV-ND, pressing in contact mode, and retracting from the quartz substrate. Simultaneously, confocal fluorescence imaging is performed to ensure that the AFM tip picks up SiV-ND. If the SiV-ND is picked up successfully, the fluorescence signal drops to the background level after the tip is retracted. The deposition procedure is started by aligning the AFM tip and the ONF to the focus point of the microscope OL. The AFM tip is adjusted such that it just touches on the surface of the ONF to deposit the SiV-ND. The whole deposition process is monitored by sending visible laser light through the ONF and observing the scattered light through the CCD. 

\begin{figure}
\centering
\includegraphics[width=9 cm]{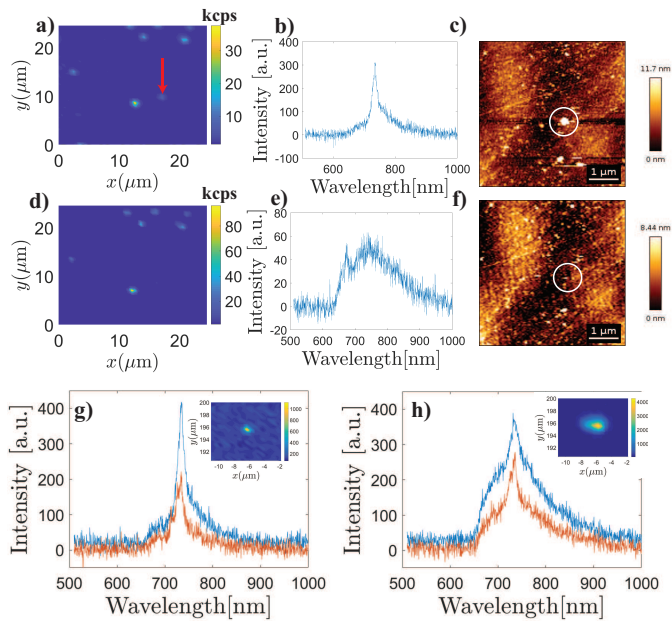}
\caption{{\label{pick and place}} (a)-(c) and (d)-(f) are fluorescence imaging and AFM topography measurements before and after picked-up the SiV-ND, respectively. (g) and (h) show the fluorescence emission spectra through the microscope OL and guided modes of the ONF, respectively. Blue and orange traces correspond to the high and low excitation laser powers, respectively. The insets show fluorescence scan images.} 
\end{figure}

A typical summary of results for the pick-up and deposition procedure is shown in Fig. \ref{pick and place}. A typical confocal fluorescence scan image is shown in Fig. \ref{pick and place}(a). The fluorescence emission spectrum of the SiV-ND for a marked position (red arrow) is shown in Fig. \ref{pick and place} (b). One can readily observe the ZPL emission from the SiV-ND.  Such a SiV-ND is located by the AFM image, as shown in Fig. \ref{pick and place} (c). The obtained SiV-ND size is $280$ ($\pm$$30$) nm. After the SiV-ND pick-up, the measured fluorescence image is shown in Fig. \ref{pick and place} (d). As one can see, fluorescence intensity was reduced to the background level of the quartz substrate. The measured spectrum at the location is shown in Fig. \ref{pick and place} (e). The excitation laser power of 56 mW (19.8 MW/$cm^2$) was used for this measurement. One can readily see that the ZPL is missing in the emission spectrum, indicating the absence of the SiV/ND. This measurement confirms that the AFM tip picks up the SiV-ND. To further ensure, the AFM imaging is performed as shown in Fig. \ref{pick and place} (f), which reveals that the SiV-ND is missing in the image. Note that SiV-NDs used in Fig.\ref{NDs pre-characterization} and Fig. \ref{pick and place} are not the same. 

After depositing on the surface of a free-standing ONF, the fluorescence emission spectrum for the same SiV-ND is measured through the microscope OL and ONF guided modes. The measured results through the microscope OL and ONF guided modes are shown in Figs. \ref{pick and place} (g) and (h), respectively. Blue and orange traces correspond to the excitation laser power of 25 mW (8.8 MW/$cm^2$) and 11 mW (3.9 MW/$cm^2$), respectively. One can readily see that emission spectra reveal peaks at the ZPL wavelength of SiV-ND, which confirms the deposition of SiV-ND on the surface of the ONF. Note that the measured spectra shown in Fig. \ref{pick and place} (h) were corrected for the ONF background. The ONF background was measured at the off-location of the SiV-ND and was subtracted from the measured spectra. The confocal fluorescence images are performed through the microscope OL and ONF guided modes for the same SiV/ND. The measured results are shown in the insets of Figs. \ref{pick and place} (g) and (h), revealing the integration of the SiV-ND on the surface of the ONF. Note that the optical transmission through the ONF dropped to $82$\% after depositing SiV-ND due to scattering.

As seen in Fig. \ref{pick and place}(h), the observed fluorescence emission spectrum through guided modes has a low signal-to-noise ratio in comparison to that measured through microscope OL, especially at high excitation power (blue trace). This is due to a lower coupling efficiency of SiV emission into ONF guided modes compared to the host ND fluorescence. In contrast, the measurements through the microscope OL are better as the confocal setup can be aligned to collect only SiV emission. We estimate the coupling efficiency into guided modes by measuring photon count rates into ONF guided modes and OL (radiation modes) as discussed in Ref.\cite{Yalla12}. To estimate absolute photon count rate into ONF guided modes and microscope OL, the light transmission efficiencies ONF to APD ($\kappa_{ONF}$) and focus point to APD ($\kappa_{OL}$) are essential. The measured $\kappa_{ONF}$ and $\kappa_{OL}$ are 25 ($\pm3$)$\%$ and 2.2 ($\pm0.4$)$\%$, respectively. The OL collection efficiency is estimated to be 23$\%$ from a numerical aperture value of 0.85. The observed background-corrected photon count rates for guided and radiation modes are 1.176$\pm 0.120$ kcps and 1.500$\pm 0.150$ kcps, respectively. The estimated coupling efficiency is 4.1$\pm 0.8$\%, which corresponds to the radial position of the dipole from the ONF surface is 110$\pm 20$ nm, assuming a random dipole-orientation\cite{Yalla12}. In the present experiments, the deposited SiV-ND size is $280$ ($\pm$$30$) nm. Therefore, the host ND fluorescence is coupled more efficiently than the SiV emission in such a case. The signal-to-noise ratio through guided modes can be improved by integrating small SiV-NDs with ONF. The fluorescence emission into ZPL from SiV-NDs can be further enhanced by incorporating a cavity structure on the ONF\cite{Yalla20, Yalla22}. The fluorescence emission spectrum from SiV-ND further can be narrowed by working at cryogenic tempeartures\cite{Neu}. 

In summary, we have demonstrated the development of advanced narrow bandwidth quantum emitters, such as SiV-NDs by employing a microwave-assisted CVD method. We have shown that SiV-NDs are picked up by the AFM tip and deposited on the surface of a free-standing ONF. The present results may lead to a new route to on-demand single-photon generation through single-mode optical fibers for potential applications in quantum information science.\\

This work was supported by the Japan Science and Technology Agency (JST) as one of the strategic innovation projects. We thank M. Morinaga for his helpful discussions.\\

The authors declare that they have no conflict of interest.\\

The data that support the findings of this study are available from the corresponding author upon reasonable request.\\


\begin{thebibliography}{} 
\newcommand{\enquote}[1]{``#1''}
\bibitem{Sipahigil} A. Sipahigil, R. E. Evans, D. D. Sukachev, M. J. Burek, J. Borregaard, M. K. Bhaskar, C. T. Nguyen, J. L. Pacheco, H. A. Atikian, C. Meuwly, R. M. Camacho, F. Jelezko, E. Bielejec, H. Park, M. Loncar, and M. D. Lukin,  Science {\bf{10}},1126-6875 (2016).
\bibitem{Lodahl} P. Lodahl, S. Mahmoodian, and S. Stobbe, Rev. Mod. Phys. \textbf{87}, 347 (2015).
\bibitem{Englund10}  D. Englund, B. Shields, K. Rivoire, F. Hatami, J. Vučković, H. Park, and M. D. Lukin, Nano Lett. {\bf{10}}, 3922-3926 (2010).
\bibitem{Riedrich}  J. Riedrich-Möller, C. Arend, C. Pauly, F. Mucklich, M. Fischer, S. Gsell, M. Schreck, and C. Becher, Nano Lett. {\bf{14}}, 5281-5287 (2014).
\bibitem{Review18} K. P. Nayak, M. Sadgrove, R. R. Yalla, F. L. Kien, and K. Hakuta, J. Opt. {\bf{20}}, 073001 (2018).
\bibitem{Kurtsiefer} C. Kurtsiefer, S. Mayer, P. Zarda, and H. Weinfurter, Phys. Rev. Lett. {\bf{85}}, 290-293 (2000).
\bibitem{Aharonovichrev}  I. Aharonovich, D. Englund, and M. Toth, Nature Photonics \textbf{10}, 631-641 (2016). 
 \bibitem{Ladd}T. D. Ladd, F. Jelezko, R. Laflamme, Y. Nakamura, C. Monroe, and J. L. O’Brien, Nature {\bf{464}}, 45-53 (2010).
\bibitem{Aharonovich11} I. Aharonovich, S. Castelletto, D. A. Simpson, C-H. Su, A. D. Greentree, and S. Prawer, Rep. Prog. Phys. \textbf{74}, 076501 (2011).
\bibitem{Rabeau} J. R. Rabeau, Y. L. Chin, S. Prawer, F. Jelezko, T. Gaebel, and J. Wrachtrup,  Appl. Phys. Lett. {\bf{86}}, 131926 (2005).
 \bibitem{Englund} T. Schröder, S. L. Mouradian, J. Zheng, M. E. Trusheim, M. Walsh, E. H. Chen, L. Li, I. Bayn, and D. Englund, J. Opt. Soc. Am. B {\bf{33}}, B65-B83 (2016). 
\bibitem{Neu} E. Neu, D. Steinmetz, J. Riedrich-Möller, S. Gsell, M. Fischer, M. Schreck, and C. Becher, New J. Phys. {\bf{13}}, 025012 (2011).
\bibitem{Meijer}J. Meijer, B. Burchard, M. Domhan,C. Wittmann, T. Gaebel, I. Popa, F. Jelezko, and J. Wrachtrup, Appl. Phys. Lett. {\bf{87}}, 261909 (2005).
\bibitem{Davydov} V. A. Davydov, A. V. Rakhmanina, S. G. Lyapin, I. D. Ilichev, K. N. Boldyrev, A. A. Shiryaev, and V. N. Agafonov, JETP Lett. {\bf{99}}, 585-9 (2014).
\bibitem{Jantzen} U. Jantzen, A. B. Kurz, D. S. Rudnicki, C. Schäfermeier, K. D. Jahnke, U. L. Andersen, V. A. Davydov, V. N. Agafonov, A. Kubanek, L. J. Rogers, and F. Jelezko, New J. Phys. {\bf{18}}, 073036 (2016).
\bibitem{Takashima} H.Takashima, A. Fukuda, K. Shimazaki, Y. Iwabata, H. Kawaguchi, A. W. Schell, T. Tashima, H. Abe, S. Onoda, T. Ohshima, and S. Takeuchi, Opt. Mater. Express {\bf{11}}, 1978-1988 (2021).
\bibitem{Yalla12} R. R. Yalla, F. L. Kien, M. Morinaga, and K. Hakuta, Phys. Rev. Lett. {\bf{109}}, 063602 (2012).
\bibitem{Yalla14}R.  R. Yalla, M. Sadgrove, K. P. Nayak, and K. Hakuta, Phys. Rev. Lett. {\bf{113}}, 143601 (2014).
\bibitem{Kali19} K. P. Nayak, J. Wang, and J. Keloth, Phys. Rev. Lett. {\bf{123}}, 213602 (2019).
\bibitem{Yalla11}R.  R. Yalla, K. P. Nayak, and K. Hakuta, Opt. Express {\bf{20}}, 2932 (2012). 
\bibitem{Schroder} T. Schröder, M. Fujiwara, T. Noda, H.-Q. Zhao, O. Benson, and S. Takeuchi, Opt. Express {\bf{20}}, 10490–10497 (2012).
\bibitem{Liebermeister} L. Liebermeister, F. Petersen, A. V. Münchow, D. Burchardt, J. Hermelbracht, T. Tashima, A. W. Schell, O. Benson, T. Meinhardt, A. Krueger, A. Stiebeiner, A. Rauschenbeutel, H. Weinfurter, and M. Weber, Appl. Phys. Lett. {\bf{104}}, 031101 (2014).
\bibitem{Schell15} A. W. Schell, H. Takashima, S. Kamioka, Y. Oe, M. Fujiwara, O. Benson, and S. Takeuchi, Sci. Rep. \textbf{5}, 9619 (2015).
\bibitem{Schell11} A. W. Schell, Gunter Kewes, T. Schroder, J. Wolters, T. Aichele and O. Benson, Rev. Sci. Instrum. {\bf{82}}, 073709 (2011).
\bibitem{Shafi} K. M. Shafi, W. Luo, R. R. Yalla, K. Iida, E. Tsutsumi, A. Miyanaga, and K. Hakuta, Sci. Rep. {\bf{8}}, 13494 (2018).
\bibitem{Yalla20}R. R. Yalla and K. Hakuta, Appl. Phys. B {\bf{126}}, 187 (2020).
\bibitem{Yalla22} R. R. Yalla, K. M. Shafi, K. P. Nayak, and K. Hakuta, Appl. Phys. Lett. {\bf{120}}, 071108 (2022).
\end{thebibliography}
\end{document}